\documentclass[runningheads]{llncs}
\usepackage{times}
\usepackage{epsfig}
\usepackage{graphicx}
\usepackage{amsmath}
\usepackage{amssymb}
\usepackage{booktabs}
\usepackage{multirow}
\usepackage{makecell}
\usepackage{xcolor}
\makeatletter 
\newcommand\figcaption{\def\@captype{figure}\caption} 
\newcommand\tabcaption{\def\@captype{table}\caption} 
\makeatother
\begin{document}
	\title{Encoding Metal Mask Projection for Metal Artifact Reduction in Computed Tomography}
	\titlerunning{Encoding Metal Mask Projection for Metal Artifact Reduction}
	%
	\author{Yuanyuan Lyu\inst{1} \and
	Wei-An Lin\inst{2} \and
	Haofu Liao\inst{3} \and
	Jingjing Lu\inst{4,5} \and 
	S. Kevin Zhou\inst{6,7}}
	\authorrunning{Y. Lyu et al.}
	\institute{Z$^2$Sky Technologies Inc., Suzhou, China \and 
	Adobe, CA, USA \and 
	Department of Computer Science, University of Rochester, NY, USA \and 
	Department of Radiology, Beijing United Family Hospital, Beijing, China \and
	Peking Union Medical College Hospital, Beijing, China \and
	Institute of computing  technology, Chinese academy of sciences, Beijing, China  \and
	Peng Cheng Laboratory, Shenzhen, China \email{s.kevin.zhou@gmail.com} }

	%
	\maketitle              

	\begin{abstract}
		
    	Metal artifact reduction (MAR) in computed tomography (CT) is a notoriously challenging task because the artifacts are structured and non-local in the image domain. However, they are inherently local in the sinogram domain. Thus, one possible approach to MAR is to exploit the latter characteristic by learning to reduce artifacts in the sinogram. However, if we directly treat the metal-affected regions in sinogram as missing and replace them with the surrogate data generated by a neural network, the artifact-reduced CT images tend to be over-smoothed and distorted since fine-grained details within the metal-affected regions are completely ignored. In this work, we provide analytical investigation to the issue and propose to address the problem by (1) retaining the metal-affected regions in sinogram and (2) replacing the binarized metal trace with the metal mask projection such that the geometry information of metal implants is encoded. Extensive experiments on simulated datasets and expert evaluations on clinical images demonstrate that our novel network yields anatomically more precise artifact-reduced images than the state-of-the-art approaches, especially when metallic objects are large.
    	\keywords{Artifact Reduction  \and Sinogram Inpainting \and Image Enhancement.}
	\end{abstract}

	\section{Introduction}
	Modern computed tomography (CT) systems are able to provide accurate images for diagnosis \cite{zhou2017deep,meyer2010normalized,zhou2015medical}. However, highly dense objects such as metallic implants cause inaccurate sinogram data in projection domain, which leads to non-local streaking artifacts in image domain after reconstruction. The artifacts degrade the image quality of CT and its diagnostic value. The challenge of metal artifact reduction (MAR) aggravates \textit{when metallic objects are large}.
	
	Conventional MAR algorithms can be grouped into three categories: iterative reconstruction, image domain MAR and sinogram domain MAR. Iterative approaches are often time-consuming and require hand-crafted regularizers, which limit their practical impacts\cite{chang2018prior,jin2015model}. Image domain methods aim to directly estimate and then remove the streak artifacts from the original contaminated image by image processing techniques~\cite{soltanian1996ct,karimi2015metal}, but they achieve limited success in suppressing artifacts. Sinogram domain methods treat metal-affected regions in sinogram as missing and replace them by interpolation~\cite{kalender1987reduction} or forward projection \cite{meyer2010normalized} but they would introduce streak artifacts tangent to the metallic objects, as the discontinuity in sinogram is hard to avoid.

	Recently, convolutional neural networks (CNNs) has been applied to solve MAR based on sinogram completion~\cite{ghani2019fast,cnnmar} or image processing~\cite{wang2018conditional}. DuDoNet~\cite{lin2019dudonet} been recently proposed to reduce the artifacts jointly in sinogram and image domains, which offers advantages over the single domain methods. Specifically, DuDoNet consists of two separate networks, one for sinogram enhancement (SE) and the other for image enhancement (IE). These two networks are connected by a Radon inversion layer (RIL) to allow gradient propagation during training. 
	
	However, there are still some limitations in DuDoNet~\cite{lin2019dudonet}. First, in the SE network, a binarized metal trace map is used to indicate the presence of metal in the sinogram. We will theoretically show that such a binarized map is a rather crude representation that \textit{totally discards} the details inside the metal mask projection. Second, in DuDoNet, the dual-domain enhancement is applied to linearly interpolated sinograms and the correspondingly reconstructed CTs. As linear interpolation only provides a rough estimate to the corrupted sinogram data, the artifact reduced images tend to be over-smoothed and severely distorted around regions with high-density materials, e.g. bones. Finally, the training data in DuDoNet are simulated by a limited number of projection angles and rays and consequently, metal artifact is compounded by strong under-sampling effect. 
	
	To address these problems of DuDoNet \cite{lin2019dudonet}, we present a novel approach utilizing the realistic information in the original sinogram and image while clearly specifying the \textit{metal mask projection}, whose importance is justified via our theoretical derivation. Furthermore, we introduce a padding scheme that is designed for sinogram and increase the number of projection angles and rays to mitigate the under-sampling effect. We boost the MAR performance of DuDoNet by a large margin (over 4dB) on a large-scale database of simulated images. The improvement is more evident when metallic objects are large. Expert evaluations confirm the efficacy of our model on clinical images too.
	
	\section{Problem Formulation}
	
	CT images represent spatial distribution of linear attenuation coefficients, which indicate the underlying anatomical structure within the human body. Let $ X(E) $ denote the linear attenuation image at energy level $ E $. According to Lambert-Beer's Law, the ideal projection data (sinogram) $ S $ detected by the CT scanner can be expressed as:
	\begin{equation} \label{eq_poly_porjction}
	S = -ln\int \eta(E)e^{-\mathcal{P}(X(E))}d E,
	\end{equation}
	where $\eta(E)$ represents fractional energy at $E$ and $ \mathcal{P} $ denotes a forward projection (FP) operator. 

	When metallic implants are present, $ X(E)$ has large variations with respect to $ E $ because mass attenuation coefficient of metal $\lambda_m(E)$ varies rapidly against $E$:
	\begin{equation} 
	X(E) = X_r + X_m(E) = X_r  + \lambda_m(E)\rho_m M,
	\end{equation}
	where $ X_m(E) $ denotes the linear attenuation image of the metallic implants, $ X_r $ denotes the residual image without the implants and is almost constant with respect to $E$, $ \rho_m $ is the density of metal, and $ M $ denotes a metal mask.  
	According the linearity of $\mathcal{P}$, the forward projection of $ X_m(E) $ can be written as:
	\begin{equation}\label{eq_metal_proj}
	\mathcal{P}(X_m(E)) = \lambda_m(E)\rho_m  \mathcal{P}(M)= \lambda_m(E)\rho_m M_{p},
	\end{equation}
	where $M_{p} = \mathcal{P}(M)$ is the \textit{metal mask projection}. 
	Substituting (\ref{eq_metal_proj}) into (\ref{eq_poly_porjction}) yields
	\begin{equation} \label{eq_sino_ma}
	S_{ma} = \mathcal{P}(X_r) -ln\int \eta(E)e^{-\lambda_m(E)\rho_m M_{p}}d E.
	\end{equation}
	Here, the first term $\mathcal{P}(X_r)$ is the projection data originated from $X_r$. The second term brings metal artifacts. Sinogram domain MAR algorithms aim to restore a clean counterpart $ S^* $ (ideally $ S^*=  \mathcal{P}(X_r)$ ) from the contaminated sinogram $ S_{ma} $. Then, an artifact-reduced image $ X^* $ could be inferred by the filtered back projection (FBP) algorithm, that is, $X^*=\mathcal{P}^{-1}(S^*) $.

	\begin{figure*}[t]
		\begin{center}
			\includegraphics[width=0.95\linewidth]{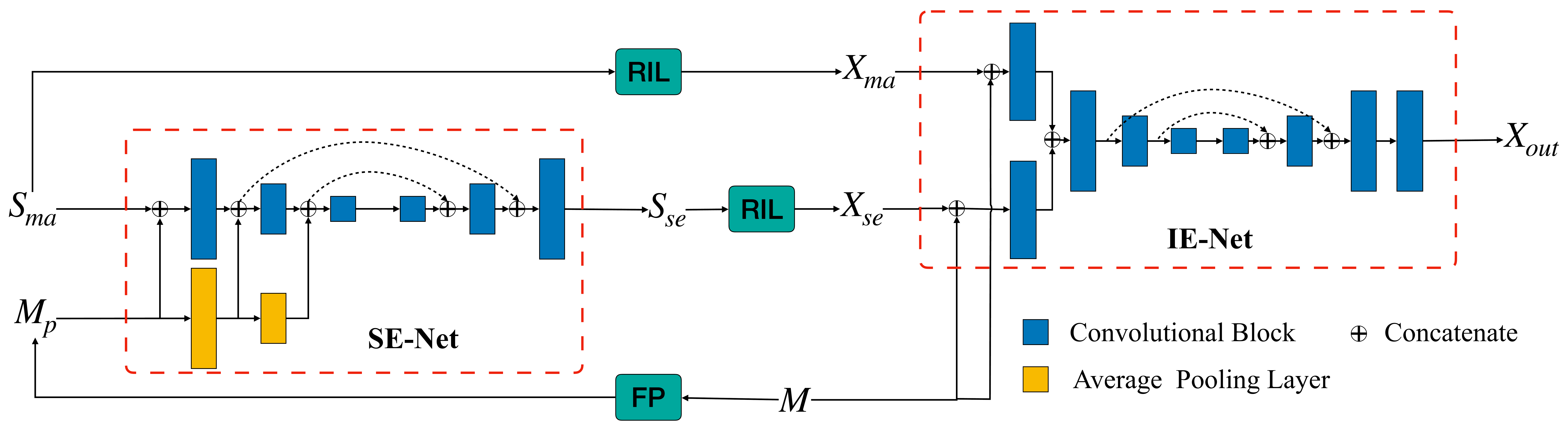}
		\end{center}
		\caption{The proposed network architecture.}
		\label{fig:network}
	\end{figure*}

	\section{Network Architecture}
	Following \cite{lin2019dudonet}, we use a sinogram enhancement network (SE-Net) and an image enhancement network (IE-Net) to jointly restore a clean image. Fig.~\ref{fig:network} shows the architecture of our proposed network. 
	
	\textbf{SE-Net.}
	To restore a clean sinogram from $S_{ma}$, conventional methods remove the second term in (\ref{eq_sino_ma}) through inpainting. Following this concept, DuDoNet takes linearly interpolated sinogram $S_{LI}$ and binarized metal trace $M_t$ as inputs for sinogram domain enhancement, where $ M_t = \delta[M_p>0]  $ ($\delta[true]=1$, $\delta[false]=0$). Here, we observe that the second term in (\ref{eq_sino_ma}) is actually a function of $M_p$. Therefore, we propose to directly \textit{utilize the knowledge of metal mask projection $ M_{p}$ }. As shown in Fig.1, our SE-Net uses a pyramid U-Net architecture $\phi_{SE}$ \cite{liao2019generative}, which takes both $ X_{ma} $ and $ M_{p} $ as inputs. To retain the projection information, $M_p$ goes through average pooling layers and then fuse with multi-scale feature maps. As metals only affect part of the sinogram data of the corresponding projection pathway, SE-Net learns to correct sinogram data within the metal trace and outputs the enhanced sinogram $S_{se}$. Sinogram enhanced image $ X_{se} $ is reconstructed by the differentiable RIL first introduced in~\cite{lin2019dudonet}, that is, $ X_{se}=\mathcal{P}^{-1}(S_{se}) $.
	
	Sinogram data is inherently periodic along the projection direction, while DuDoNet uses zero padding for convolutions in SE-Net which ignores the periodic information. Here, to offer more useful information for convolution,  we propose a new padding strategy for sinogram data using periodic padding along the direction of projection angles and zero padding along the direction of detectors, as shown in Fig.~\ref{fig:sino_pad}. 
	
	\textbf{IE-Net.}
	To suppress the secondary artifacts in $X_{se}$, we apply an image enhancement net, which refines $X_{se}$ with $M$ and $X_{ma}$. The network contains two initial convolutional layers, a U-net \cite{ronneberger2015u} and a final convolutional layer. To pay attention to the strongly distorted regions, we concatenate an image ($ X_{se} $ or $ X_{ma} $) with metal mask $M$ separately and obtain mask-aware feature maps by an initial convolutional layer with 64 3 $ \times $ 3 kernels. The two sets of mask-aware feature maps are concatenated as the input for the subsequent U-Net. A U-Net of depth 4 is used which outputs a feature map with 64 channels. Finally, a convolutional layer is used as the output layer which generates the enhanced image $X_{out}$.
	
	\textbf{Learning.}
	The total loss of our model consists of sinogam enhancement loss, image enhancement loss and Radon consistency loss~\cite{lin2019dudonet}:
	\begin{equation}
	\mathcal{L} _{total}= \alpha_{se} || S_{se} - S_{gt}||_1 +(\alpha_{rc} ||X_{se} - X_{gt}||_1 + \alpha_{ie} ||X_{out} - X_{gt}||_1 )\odot (1 - M),
	\end{equation}
	where $\alpha_{se}$ , $\alpha_{rc}$, and $\alpha_{ie}$ are blending weights. We empirically set them to 1.
	
	\section{Experiment}
	
	\subsection{Dataset and Experimental Setup}
\begin{figure*}[h]
		\begin{center}
			\includegraphics[width=0.9\linewidth]{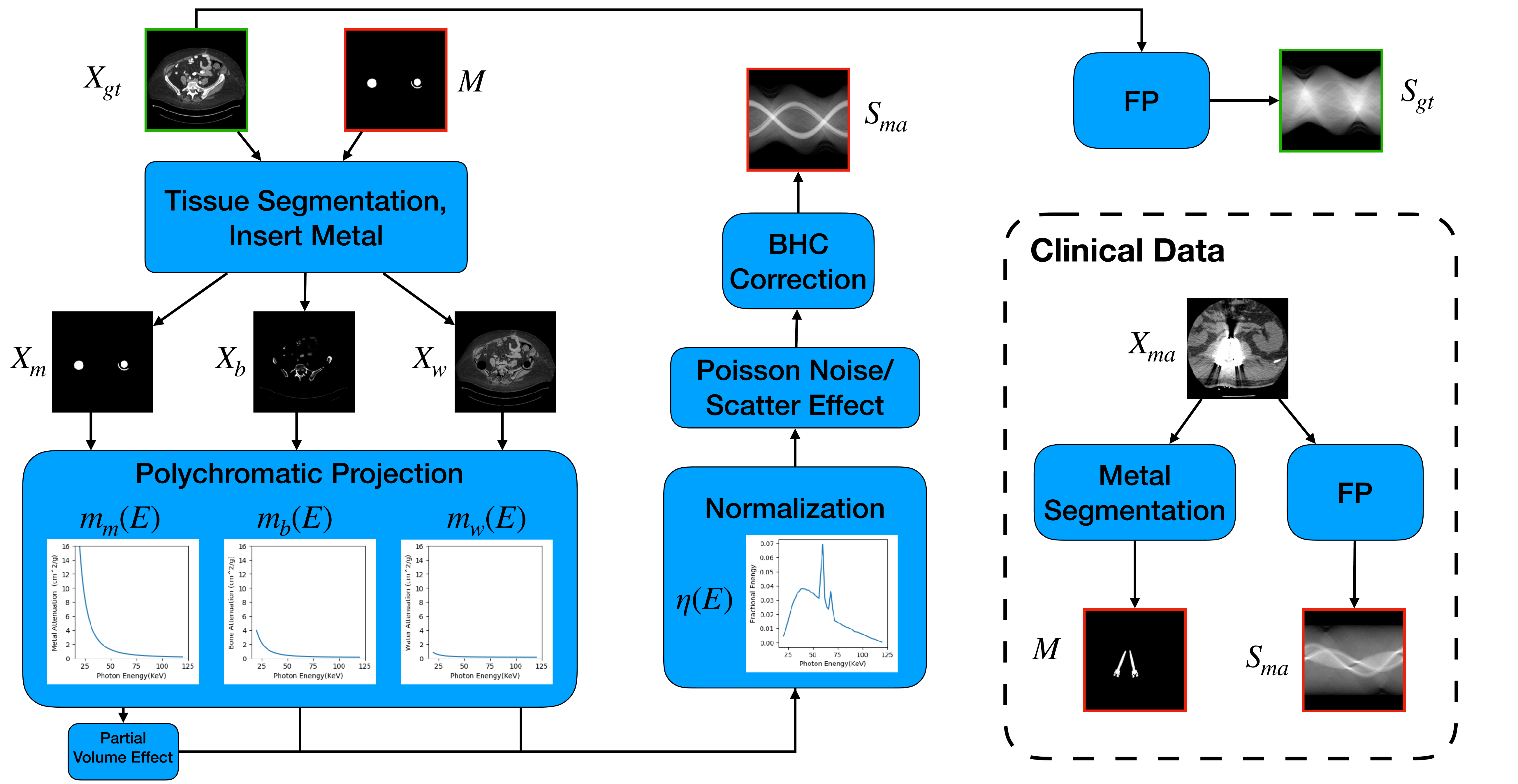}
		\end{center}
		\caption{Flowchart of metal artifact simulation and data generation of clinical images. Images with red borders are the inputs of our model and images with green borders are ground truth.}
		\label{fig:data}
	\end{figure*}

	\textbf{Simulation data.}
	We generate 360,000 cases for training and 2,000 cases for testing based on clean CT images. We first resize CT images to a  size of 416$\times$416 and use 640 projection angles and 641 rays for imaging geometry to simulate realistic metal artifacts (details are presented in Fig.~\ref{fig:data}).  
		
	\noindent\textbf{Clinical data.}
	We evaluate the proposed method using two clinical datasets. We refer them to DL and CL. DL represents the DeepLesion dataset~\cite{yan2018deep} and CL is a clinical CT scan for a patient with metal rods and screws after spinal fusion. We randomly select 30 slices from DL and 10 slices from CL with \textit{more than 100 pixels above 3,000 HU and moderate or severe metal artifacts}. The clinical images are resized and processed with the same geometry as the simulation data (see Fig.~\ref{fig:data}).
	
	\noindent\textbf{Implementation and training details.}
	Our model is implemented using the PyTorch framework. We use the Adam optimizer with $ (\beta_1, \beta_2) = (0.5, 0.999) $ to train the model. The learning rate starts from 0.0002 and is halved for every 30 epochs. The model is trained on an Nvidia 2080Ti GPU card for 201 epochs with a batch size of 2.
	
	\noindent\textbf{Metrics.}
	We use peak signal-to-noise ratio (PSNR) and structural similarity index (SSIM) to evaluate the corrected image with a soft tissue window in the range of [-175, +275] HU. To evaluate the sinogram restoration performance, we use mean square error (MSE) to compare the enhanced $S_{se}$ with $S_{gt}$. We group results according to the size of metal implants to investigate the MAR performance. 
	
	\noindent\textbf{Rating.}
	A proficient radiologist with about 20 years of reading experience is invited to rate the image quality for each group of the corrected images by paying close attention to ameliorating beam hardening, removing primary streaky artifact, reducing secondary streaky artifacts and overall image quality. The radiologist is asked to rate all the images from each group in a random order, with a rating from 1, indicating very good MAR performance, to 4, not effective at all. We use paired T-test to compare the ratings between our model and every state-of-the-art method. 
	
	\subsection{Ablation Study}
	In this section, we investigate the effectiveness of different modules of the proposed architecture. 
	We use the following configurations for this ablation study:
	
	\renewcommand{\labelenumi}{\alph{enumi})}
	\begin{enumerate}
		\item IE-Net: the IE network with $X_{ma}$ and $M$,
		\item SE$_0$-Net: the SE network with $S_{ma}$ and $M_t$,
		\item SE-Net: the SE network with $S_{ma}$ and $M_p$,
		\item SE$_p$-Net: the SE-Net with sinogram padding,
		\item SE$_p$-IE-Net: the SE$_p$-Net with an IE-Net to refine $X_{se}$ with $M$,
		\item Ours: our full model, SE$_p$-IE-Net refined with $X_{ma}$.
	\end{enumerate}
	
	\begin{table*} [t]
		\footnotesize
		\begin{center}
			\resizebox{\linewidth}{!}{
				\begin{tabular}{l|ccccc|c}
					\toprule[1pt]
					&&Large Metal& $\rightarrow$ &Small Metal&& Average\\
					\hline
					\hline
					$X_{ma}$&19.42/81.1/1.1e+1&23.07/85.4/7.3e+0&26.12/88.7/2.2e+0&26.60/89.3/1.7e+0&27.69/89.9/3.8e-1&24.58/86.9/4.5e+0\\
					IE-Net&31.19/94.8/ $~~$n.a.$~~~$ &30.33/95.9/ $~~$n.a.$~~~$ &34.48/96.8/ $~~$n.a.$~~~$ &35.52/96.8/ $~~$n.a.$~~~$ &36.37/97.0/ $~~$n.a.$~~~$ &33.58/96.3/ $~~$n.a.$~~~$ \\
					SE$_0$-Net&20.28/86.5/3.0e+0&21.65/89.6/1.6e+0&26.39/91.7/3.0e-2&25.65/91.3/6.4e-2&24.93/91.1/8.4e-2&23.78/90.0/9.5e-1\\
					SE-Net&26.71/91.0/2.7e-3&27.93/92.6/4.3e-4&28.20/93.2/2.4e-4&28.31/93.2/1.8e-4&28.34/93.3/\textbf{1.4e-4}&27.90/92.7/7.4e-4\\
					SE$_p$-Net&26.86/91.0/2.2e-3&27.94/92.5/4.4e-4&28.20/93.1/2.4e-4&28.31/93.2/1.9e-4&28.34/93.3/1.7e-4&27.93/92.6/6.5e-4\\
					SE$_p$-IE-Net&34.35/96.1/\textbf{1.7e-3}&36.03/96.8/4.4e-4&37.02/97.1/2.4e-4&37.53/97.2/1.9e-4&37.64/97.3/1.5e-4&36.52/96.9/\textbf{5.5e-4}\\
					Ours&\textbf{34.60}/\textbf{96.2}/3.4e-3&\textbf{36.84}/\textbf{97.0}/\textbf{4.2e-4}&\textbf{37.84}/\textbf{97.4}/\textbf{2.2e-4}&\textbf{38.34}/\textbf{97.4}/\textbf{1.7e-4}&\textbf{38.38}/\textbf{97.5}/1.5e-4&\textbf{37.20}/\textbf{97.1}/8.8e-4\\
					\toprule[1pt]
				\end{tabular}
			}
		\end{center}
		\caption{Quantitative evaluation (PSNR(dB)/SSIM\%/MSE) for different models.}
		\label{table:ablation}
	\end{table*}
	
	\begin{figure}[h]
		\begin{minipage}[b]{0.6\textwidth}
			\centering
			\small
			\begin{minipage}[t]{0.18\textwidth}
				\centering
				\includegraphics[width=1\textwidth]{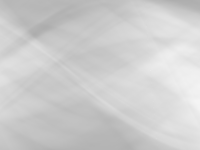}
				\includegraphics[width=1\textwidth]{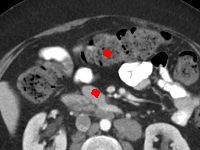}
				\includegraphics[width=1\textwidth]{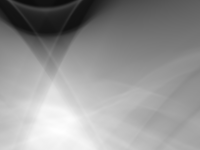}
				\includegraphics[width=1\textwidth]{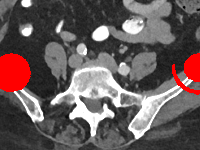}
				$S_{gt}$/$X_{gt}$
			\end{minipage}
			\begin{minipage}[t]{0.18\textwidth}
				\centering
				\includegraphics[width=1\textwidth]{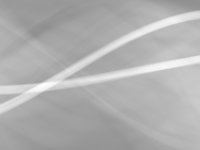}
				\includegraphics[width=1\textwidth]{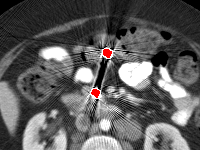}
				\includegraphics[width=1\textwidth]{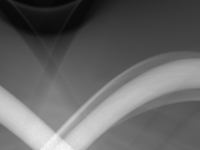}
				\includegraphics[width=1\textwidth]{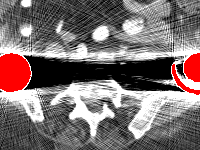}
				$S_{ma}$/$X_{ma}$
			\end{minipage}
			\begin{minipage}[t]{0.18\textwidth}
				\centering
				\includegraphics[width=1\textwidth]{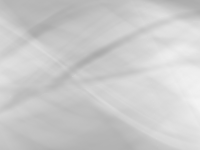}
				\includegraphics[width=1\textwidth]{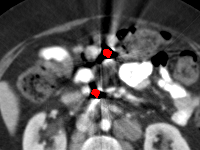}
				\includegraphics[width=1\textwidth]{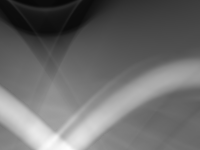}
				\includegraphics[width=1\textwidth]{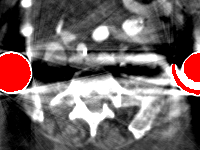}
				SE$_0$-Net
			\end{minipage}
			\begin{minipage}[t]{0.18\textwidth}
				\centering
				\includegraphics[width=1\textwidth]{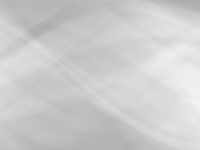}
				\includegraphics[width=1\textwidth]{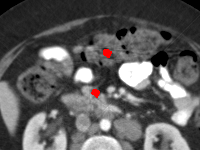}
				\includegraphics[width=1\textwidth]{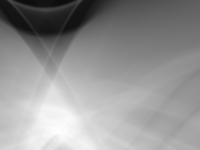}
				\includegraphics[width=1\textwidth]{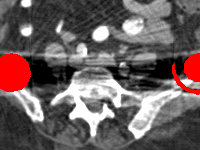}
				SE-Net
			\end{minipage}
			\begin{minipage}[t]{0.18\textwidth}
				\centering
				\includegraphics[width=1\textwidth]{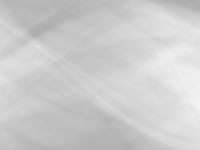}
				\includegraphics[width=1\textwidth]{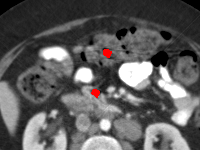}
				\includegraphics[width=1\textwidth]{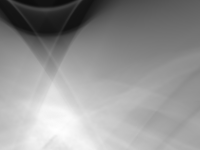}
				\includegraphics[width=1\textwidth]{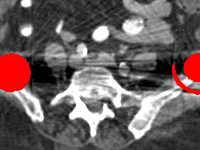}
				SE$_p$-Net
			\end{minipage}
			\caption{Comparison of different sinogram enhancement networks. The enhanced sinograms and paired CT images are presented. The red pixels stand for metal implants.}
			\label{fig:se_net}
		\end{minipage}
		\begin{minipage}[b]{0.36\textwidth}
			\centering
			\small
			\begin{minipage}[t]{0.45\textwidth}
				\centering
				\includegraphics[width=1\textwidth]{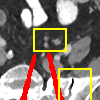}
				$X_{gt}$
				\includegraphics[width=1\textwidth]{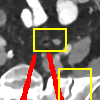}
				SE$_p$-IE-Net
			\end{minipage}
			\begin{minipage}[t]{0.45\textwidth}
				\centering
				\includegraphics[width=1\textwidth]{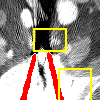}
				$X_{ma}$
				\includegraphics[width=1\textwidth]{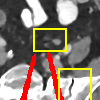}
				Ours
			\end{minipage}
			\caption{Comparison of refinement with and without $X_{ma}$. }
			\label{fig:refine_wma}
		\end{minipage}
	\end{figure}

	\noindent\textbf{Effect of metal mask projection (SE$_0$-Net vs SE-Net).} From Table \ref{table:ablation}, we can observe the use of $M_p$ instead of $M_{t}$ improves the performance for at least 4.1 dB in PNSR and reduces MSE from 0.95219 to 0.00074 for all metal sizes. The groups with large metal implants benefit more than groups with small metal implants. As shown in Fig. \ref{fig:se_net}, the artifacts in metal trace of SE$_0$-Net are over-removed or under-removed, which introduces bright and dark bands in the reconstructed CT image. With the help of $M_p$, SE-Net can suppress the artifacts even when the metallic implants are large and the surrogate data are more consistent with the correct data outside the metal trace.
	
	\noindent\textbf{Effect of sinogram padding (SE-Net vs SE$_p$-Net).}
	Sinogram padding mainly improves the performance in the group with the largest metal objects, with a PSNR gain of 0.15 dB and an MSE reduction of 0.00048. As shown in Fig. \ref{fig:se_net}, the model with sinogram padding restores finer details of soft tissue between large metallic objects because more correct information is retained by periodic padding than zero-padding.  
	
	\noindent\textbf{Effect of learning with $X_{ma}$ (SE$_p$-IE-Net vs Ours).}
	When $X_{se}$ is jointly restored with the corrupted $X_{ma}$, the sinogram correction performance is affected with an increment of 0.00033 in MSE and of 0.7 dB in PSNR. More details of soft tissue around metal are retained and the image becomes sharper, as shown in Fig. \ref{fig:refine_wma}. 
	
	\subsection{Comparison on Simulation Data}
	We compare our model with multiple state-of-the-art MAR methods. LI \cite{kalender1987reduction} and NMAR \cite{meyer2010normalized} are traditional algorithms, in which we use the simulated $S_{ma}$ as inputs. Wang \textit{et al.}~\cite{wang2018conditional} propose conditional GAN for MAR purely in image domain. Here we refer their method as cGan-CT and retrain the model using pix2pix~\cite{isola2017image} on our simulation data. For CNNMAR, we use the trained model provided by~\cite{cnnmar}. Note that DuDoNet reported here is trained on new simulation data with larger sinogram resolution (641 $\times$ 640), which is different from the sinogram resolution (321 $\times$ 320) used in \cite{lin2019dudonet}. 
	
	\begin{table*}[t]
		\centering
		\footnotesize
		\begin{center}
			\resizebox{\linewidth}{!}{
				\begin{tabular}{l|ccccc|c}
					\toprule[1pt]
					Matrics&&Large Metal& $\rightarrow$ &Small Metal&& Average\\
					\hline
					\hline
					$X_{ma}$&19.42/81.1/1.1e+1&23.07/85.4/7.3e+0&26.12/88.7/2.2e+0&26.60/89.3/1.7e+0&27.69/89.9/3.8e-1&24.58/86.9/4.5e+0\\
					cGAN-CT~\cite{wang2018conditional}&16.89/80.7/ ~~n.a.~~~ &18.35/83.7/ ~~n.a.~~~ &19.94/86.6/ ~~n.a.~~~ &21.43/87.6/ ~~n.a.~~~ &24.53/89.0/ ~~n.a.~~~ &20.23/85.5/ ~~n.a.~~~ \\
					LI~\cite{kalender1987reduction}&20.10/86.7/1.4e-1&22.04/88.7/9.4e-2&25.50/90.2/2.1e-2&26.54/90.7/1.9e-2&27.25/91.2/9.7e-3&24.28/89.5/5.7e-2\\
					NMAR~\cite{meyer2010normalized}&20.89/86.6/2.3e-1&23.73/89.7/1.3e-1&26.80/91.4/2.7e-2&27.25/91.8/3.6e-2&28.08/92.1/2.2e-2&25.35/90.3/9.0e-2\\
					CNNMAR~\cite{cnnmar}&23.72/90.1/\underline{4.4e-2}&25.78/91.6/\underline{2.4e-2}&28.25/92.6/\underline{4.7e-3}&28.87/92.9/\underline{3.3e-3}&29.16/93.1/2.0e-3&27.16/92.0/\underline{1.6e-2}\\
					DuDoNet~\cite{lin2019dudonet}&\underline{28.98}/\underline{94.5}/5.1e-2&\underline{31.00}/\underline{95.6}/3.9e-2&\underline{33.80}/\underline{96.5}/5.9e-3&\underline{35.61}/\underline{96.8}/3.6e-3&\underline{35.67}/\underline{96.9}/\underline{2.0e-3}&\underline{33.01}/\underline{96.0}/2.0e-2\\
					\hline
					Ours&\textbf{34.60}/\textbf{96.2}/\textbf{3.4e-3}&\textbf{36.84}/\textbf{97.0}/\textbf{4.2e-4}&\textbf{37.84}/\textbf{97.4}/\textbf{2.2e-4}&\textbf{38.34}/\textbf{97.4}/\textbf{1.7e-4}&\textbf{38.38}/\textbf{97.5}/\textbf{1.5e-4}&\textbf{37.20}/\textbf{97.1}/\textbf{8.8e-4}\\
					\toprule[1pt]
				\end{tabular}
			}
		\end{center}
		\caption{Quantitative evaluation for proposed network and the state-of-the-arts methods.}
		\label{table:comparison_psnr}
	\end{table*}

	\begin{figure*}[h]
		\small
		\centering
		\begin{minipage}[t]{0.11\textwidth}
			\centering
			\includegraphics[width=1\textwidth]{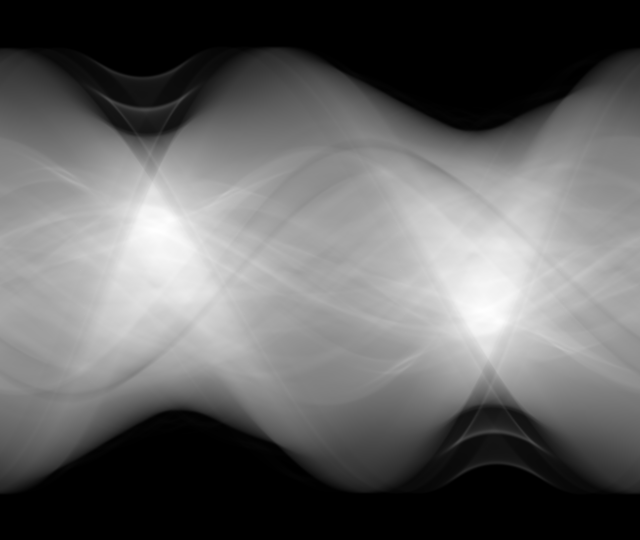}
			\includegraphics[width=1\textwidth]{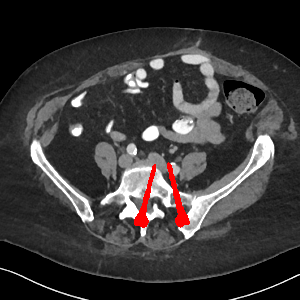}
			\includegraphics[width=1\textwidth]{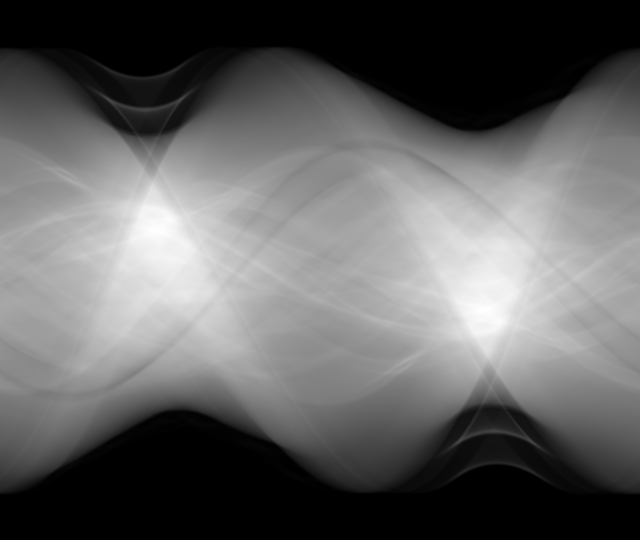}
			\includegraphics[width=1\textwidth]{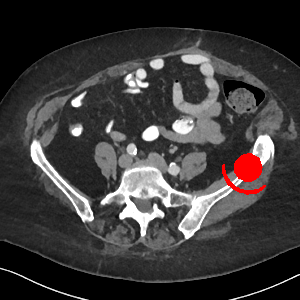}
			\scriptsize{$S_{gt}/X_{gt}$}
		\end{minipage}
		\begin{minipage}[t]{0.11\textwidth}
			\centering
			\includegraphics[width=1\textwidth]{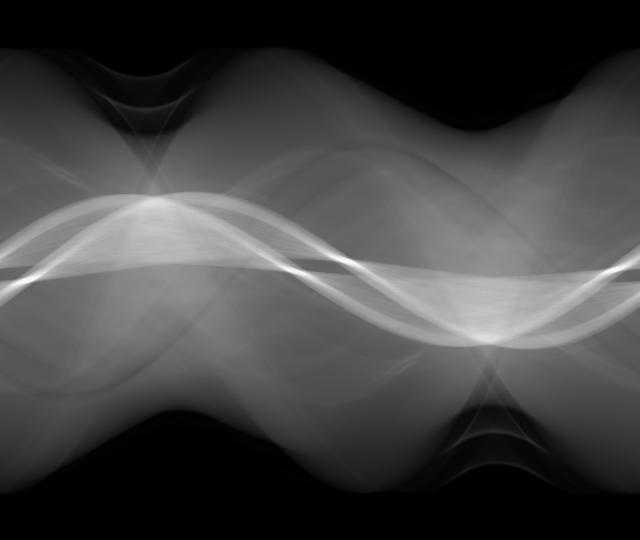}
			\includegraphics[width=1\textwidth]{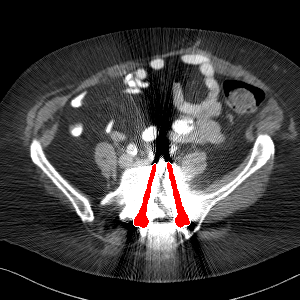}
			\includegraphics[width=1\textwidth]{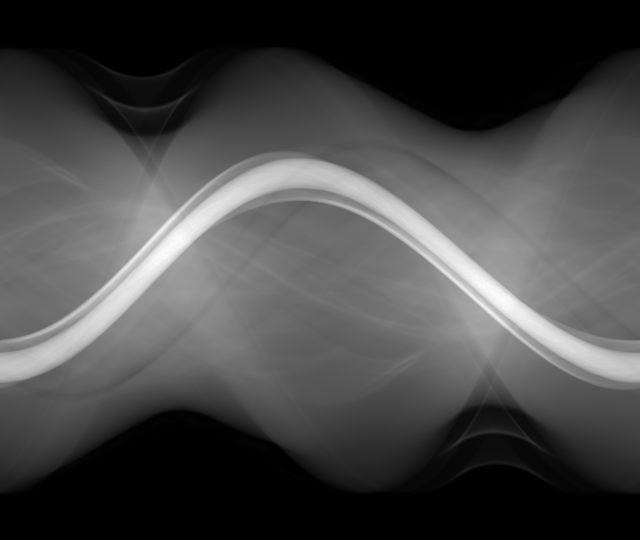}
			\includegraphics[width=1\textwidth]{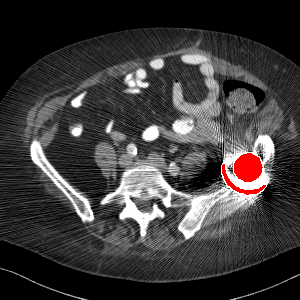}
			\scriptsize{$S_{ma}/X_{ma}$}
		\end{minipage}
		\begin{minipage}[t]{0.11\textwidth}
			\centering
			\includegraphics[width=1\textwidth]{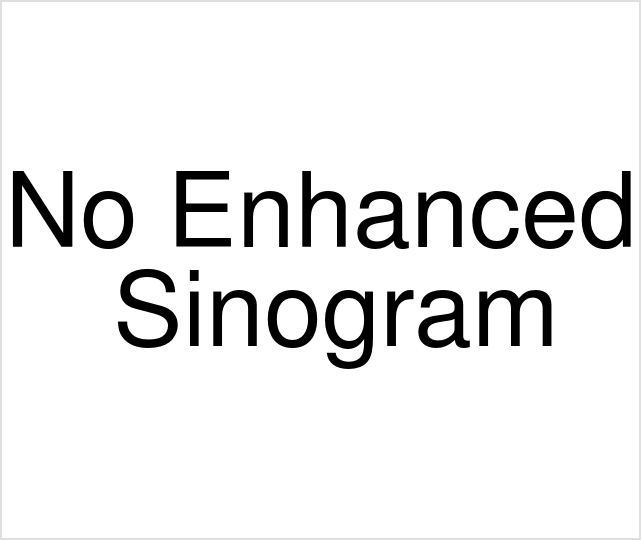}
			\includegraphics[width=1\textwidth]{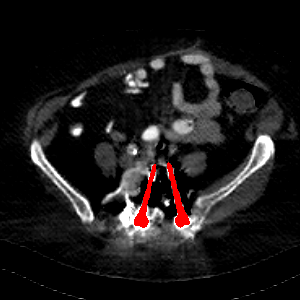}
			\includegraphics[width=1\textwidth]{figures/method_comparisons/white.png}
			\includegraphics[width=1\textwidth]{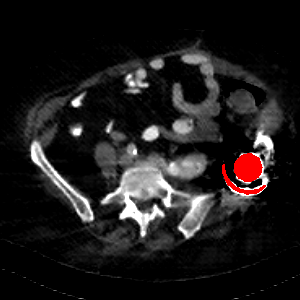}
			\scriptsize{cGAN-CT}
		\end{minipage}
		\begin{minipage}[t]{0.11\textwidth}
			\centering
			\includegraphics[width=1\textwidth]{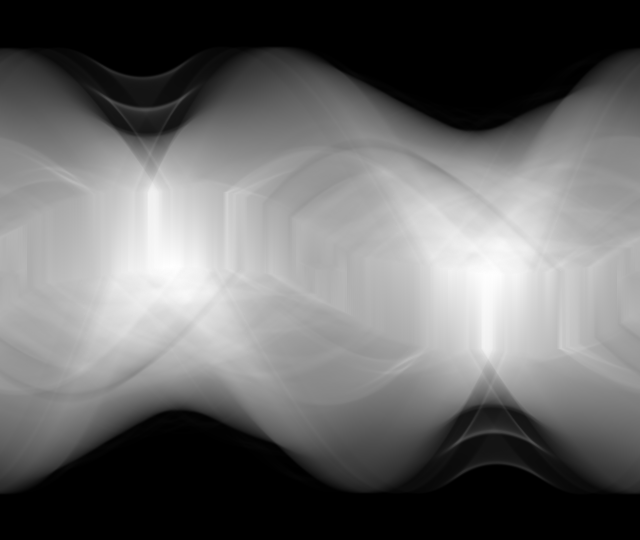}
			\includegraphics[width=1\textwidth]{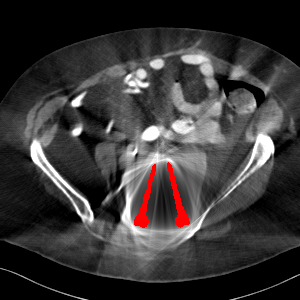}
			\includegraphics[width=1\textwidth]{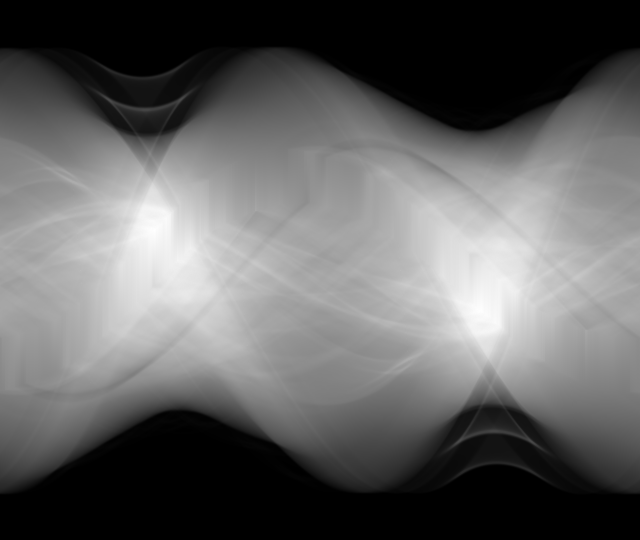}
			\includegraphics[width=1\textwidth]{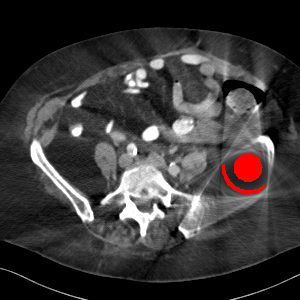}
			\scriptsize{LI}
		\end{minipage}
		\begin{minipage}[t]{0.11\textwidth}
			\centering
			\includegraphics[width=1\textwidth]{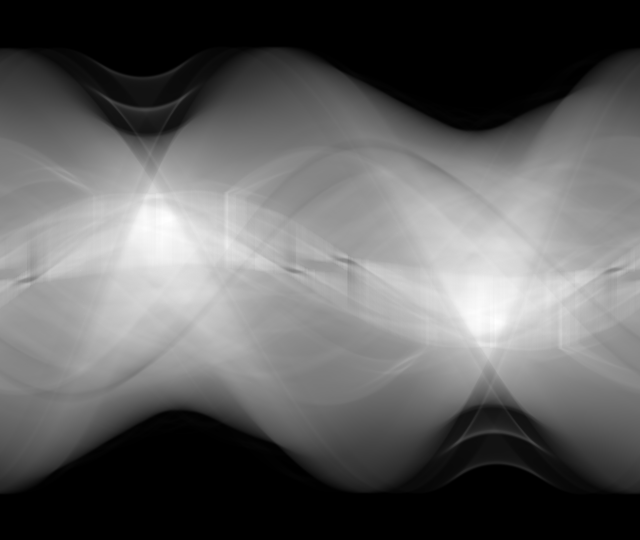}
			\includegraphics[width=1\textwidth]{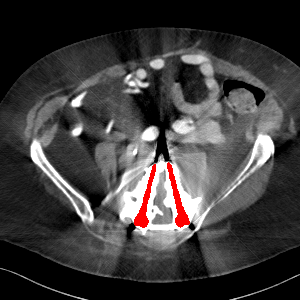}
			\includegraphics[width=1\textwidth]{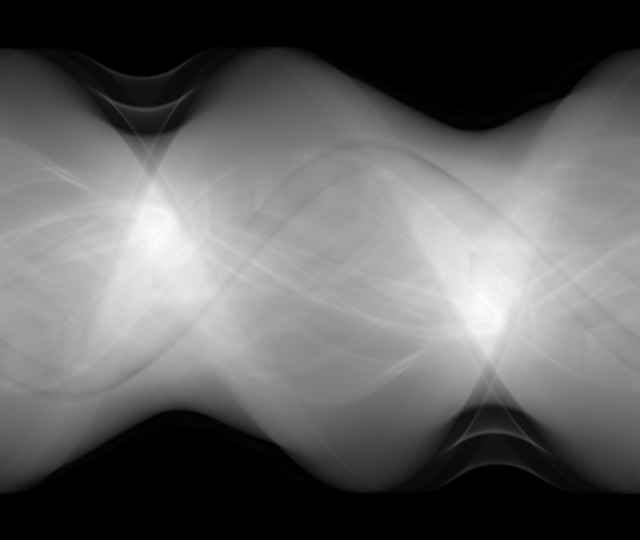}
			\includegraphics[width=1\textwidth]{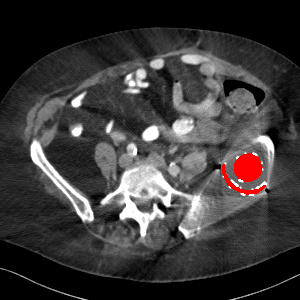}
			\scriptsize{NMAR}
		\end{minipage}
		\begin{minipage}[t]{0.11\textwidth}
			\centering
			\includegraphics[width=1\textwidth]{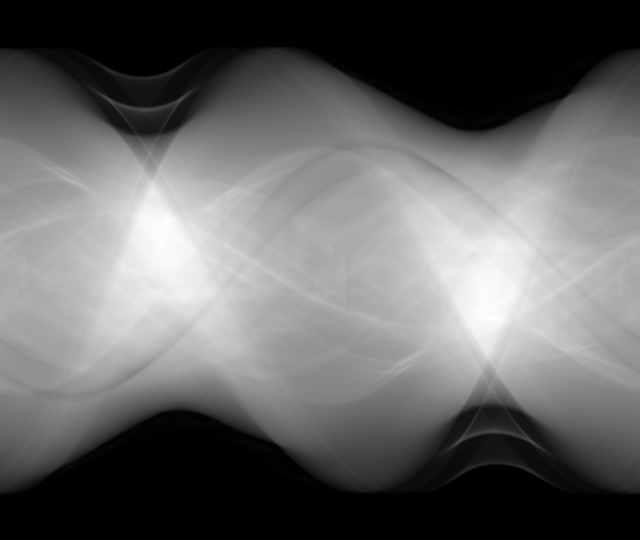}
			\includegraphics[width=1\textwidth]{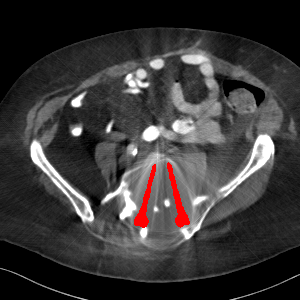}
			\includegraphics[width=1\textwidth]{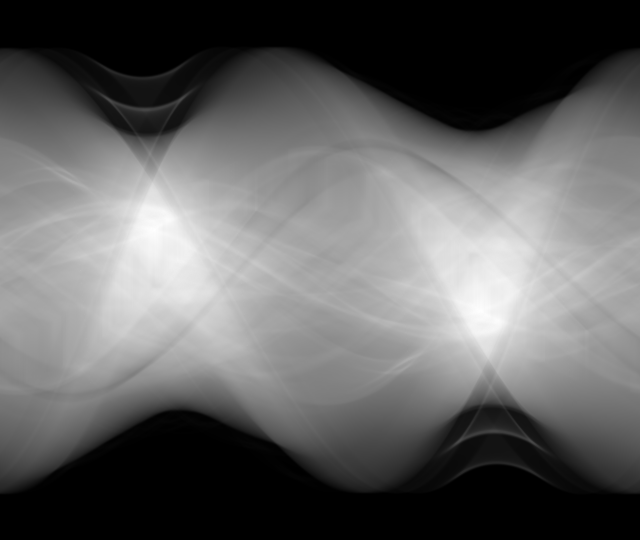}
			\includegraphics[width=1\textwidth]{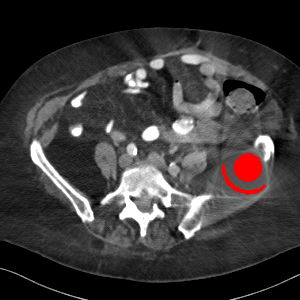}
			\scriptsize{CNNMAR}
		\end{minipage}
		\begin{minipage}[t]{0.11\textwidth}
			\centering
			\includegraphics[width=1\textwidth]{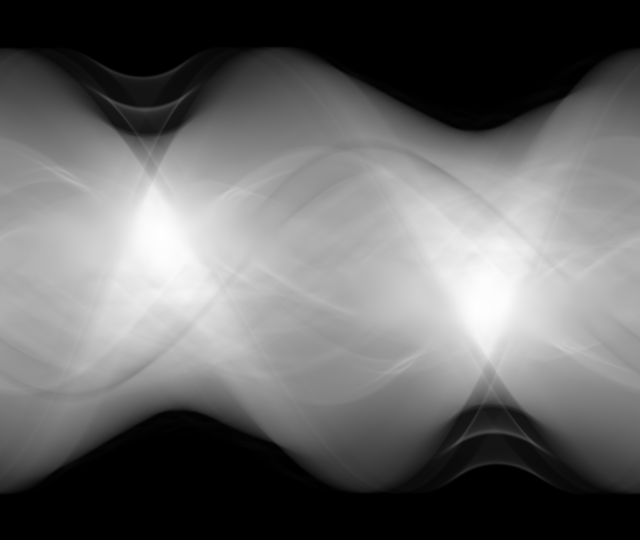}
			\includegraphics[width=1\textwidth]{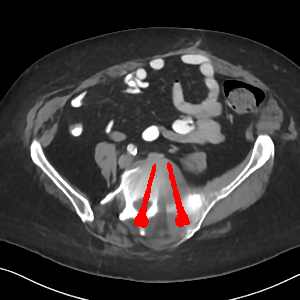}
			\includegraphics[width=1\textwidth]{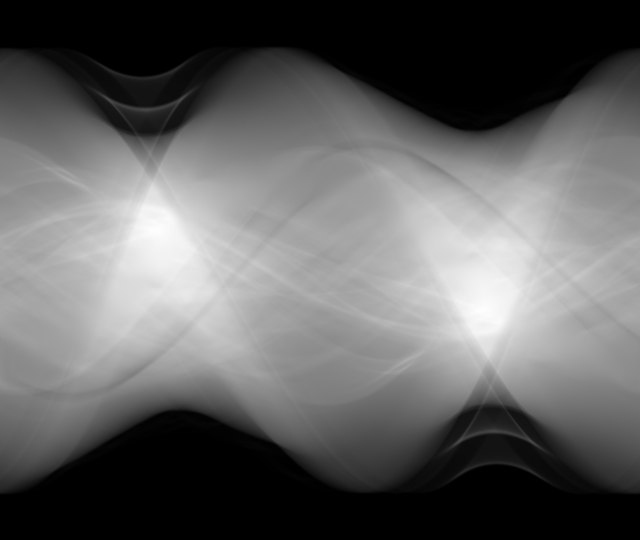}
			\includegraphics[width=1\textwidth]{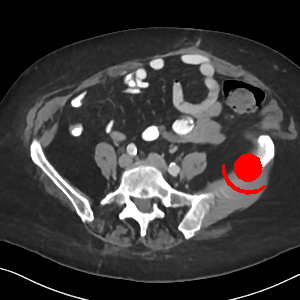}
			\scriptsize{DuDoNet}
		\end{minipage}
		\begin{minipage}[t]{0.11\textwidth}
			\centering
			\includegraphics[width=1\textwidth]{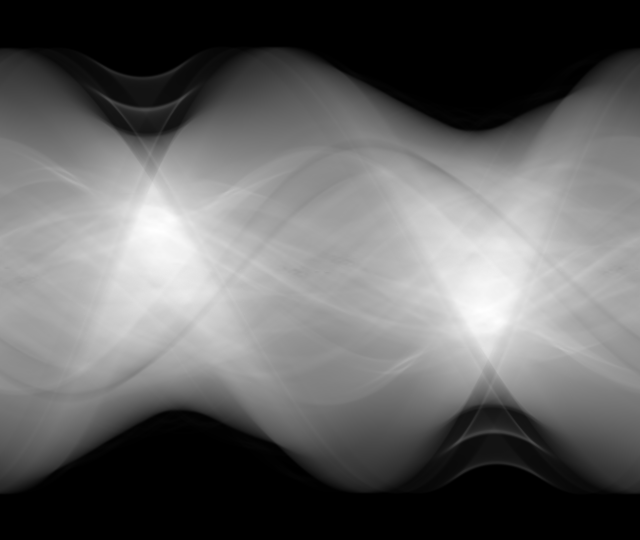}
			\includegraphics[width=1\textwidth]{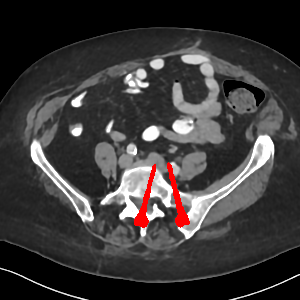}
			\includegraphics[width=1\textwidth]{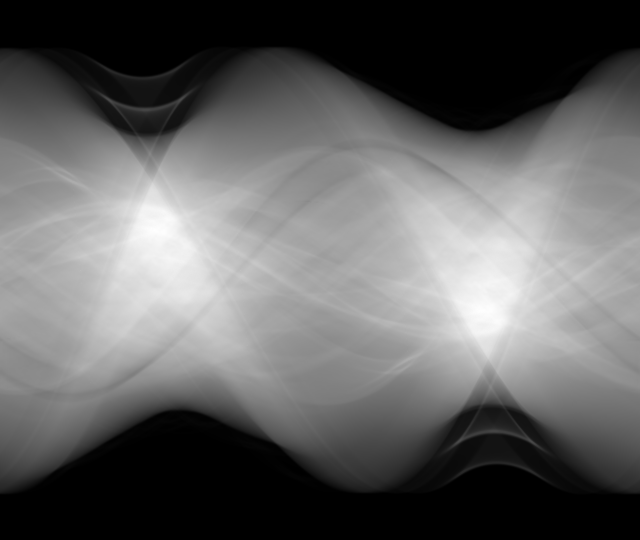}
			\includegraphics[width=1\textwidth]{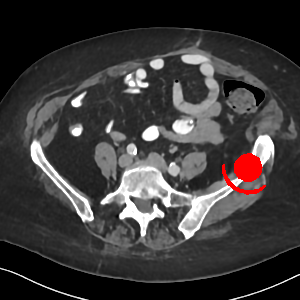}
			\scriptsize{Ours}
		\end{minipage}
		\caption{Comparison with the state-of-the-art methods on simulation data.}
		\label{fig:eval_comparison}
	\end{figure*}

	\noindent\textbf{Quantitative comparison.} 
	As shown in Table~\ref{table:comparison_psnr}, we can see all the sinogram domain MAR algorithms outperform image enhancement approach cGAN-CT in PSNR and SSIM. It is because the sinogram restoration only happens inside the metal trace and the correct sinogram data outside the metal trace help to retain the anatomical structure. CNN-based methods (CNNMAR, DuDoNet, Ours) achieve much better performance than traditional methods, with higher PSNRs and SSIMs in image domain and lower MSEs in sinogram domain. Among all the state-of-the-art methods, CNNMAR achieves the best performance in sinogram enhancement and DuDoNet achieves the best performance in reconstructed images. The proposed method attains the best performance in all metal sizes, with an overall improvement of 4.2 dB in PSNR compared with DuDoNet and 99.4\% reduction in MSE compared with CNNMAR.

	\noindent\textbf{Visual comparison.}
	As shown in Fig.~\ref{fig:eval_comparison}, metallic implants such as spinal rods and hip prosthesis cause severe streaky artifacts and metal shadows, which obscure bone structures around them. cGan-CT cannot recover image intensity correctly for both cases. Sinogram domain or dual-domain methods perform much better than cGan-CT. LI, NMAR, and CNNMAR introduce strong secondary artifacts and distort the whole images. In NMAR images, there are fake bone structures around the metals, which is related to segmentation error in the prior image from strong metal artifacts. The segmentation error is also visible in NMAR sinogram. CNNMAR cannot restore the correct bone structures between rods in case 1. The tissues around the metals are over-smoothed in DuDoNet because LI sinogram and image are used as inputs, and the missing information cannot be inferred later. Our model retains more structural information than DuDoNet and generates anatomically more faithful artifact-reduced images.

	\begin{figure}[h]
		\begin{minipage}[b]{0.4\textwidth}
			\begin{minipage}[b]{0.8\textwidth}
				\includegraphics[width=1\linewidth]{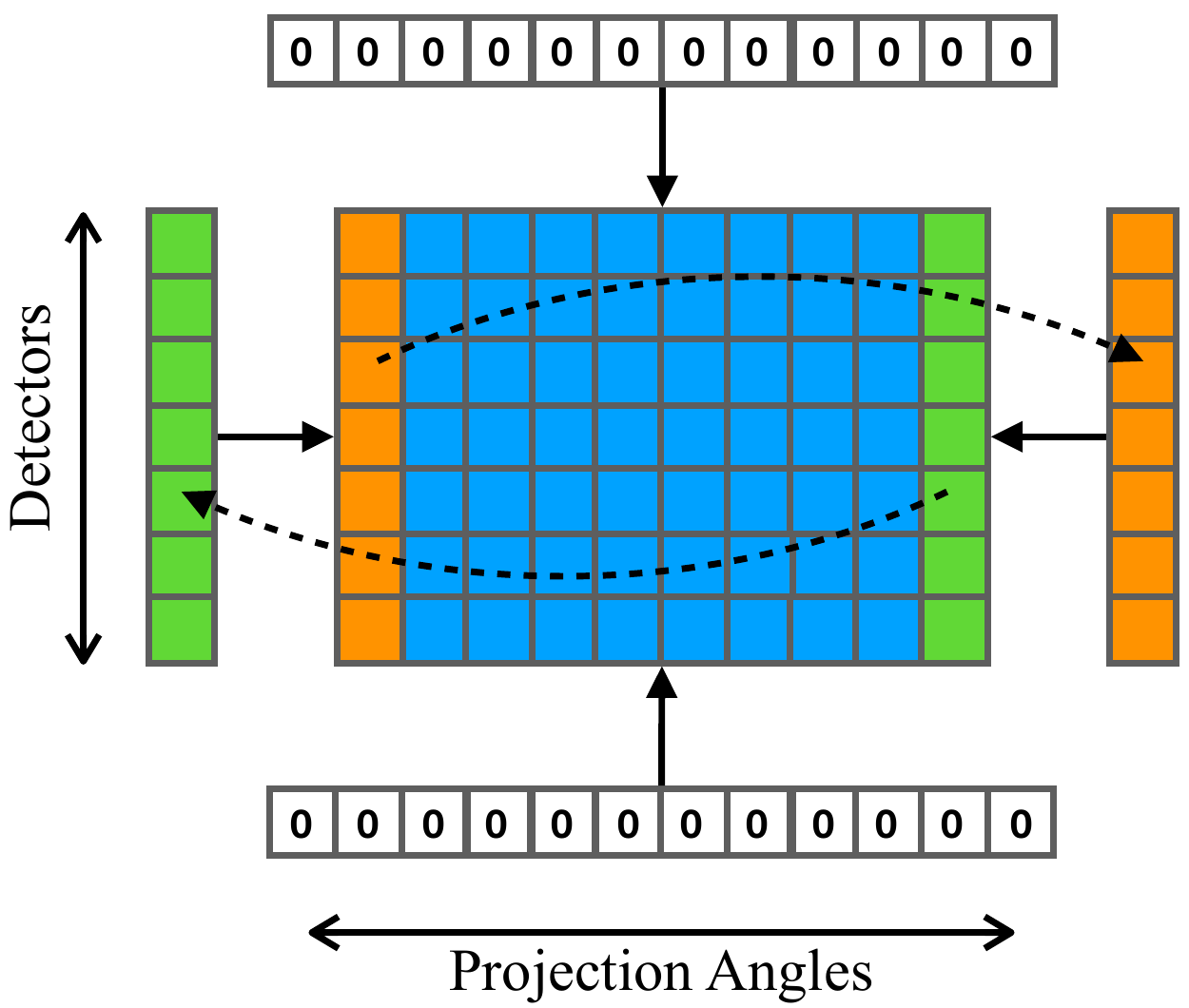}
			\end{minipage}
			\caption{Sinogram padding. }
			\label{fig:sino_pad}
		\end{minipage}
		\begin{minipage}[b]{0.6\textwidth}
			\resizebox{\linewidth}{!}{
				\begin{tabular}{l|cc|cc}
					\toprule[1pt]
					&\multicolumn{2}{c|}{DL}     &\multicolumn{2}{c}{CL} \\
					&Rating        &P Value        &Rating            &P Value\\
					\hline
					\hline
					cGAN-CT~\cite{wang2018conditional}        &2.50$\pm$0.17    &$<$0.001    &4.00$\pm$0.00    &$<$0.001        \\
					LI~\cite{kalender1987reduction}            &3.80$\pm$0.09    &$<$0.001    &3.70$\pm$0.15    &$<$0.001        \\
					NMAR~\cite{meyer2010normalized}            &2.73$\pm$0.13    &$<$0.001    &2.70$\pm$0.15    &$<$0.001        \\
					CNNMAR~\cite{cnnmar}                    &2.40$\pm$0.12    &$<$0.001    &2.20$\pm$0.20    &0.003            \\
					DuDoNet~\cite{lin2019dudonet}           &1.46$\pm$0.11    &0.312        &1.70$\pm$0.21    &0.278    \\
					\hline
					Ours                                &\textbf{1.27$\pm$0.13}    &n.a.         &\textbf{1.40$\pm$0.16}    &n.a.        \\
					\toprule[1pt]
				\end{tabular}
			}
			\tabcaption{Ratings of clinical CT images.}
			\label{table:rating}
		\end{minipage}
	\end{figure}
	
	\subsection{Clinical Study}
	\noindent\textbf{Rating.} Table \ref{table:rating} summarizes the ratings and P values for comparison between our model and the other methods. The performance of our model is significantly better than cGan-CT, LI, NMAR, CNNMAR on both datasets (all P values $\leq$ 0.03). Our model also achieves better ratings than DuDoNet.
	
\begin{figure*}[h]
		\small
		\centering
		\begin{minipage}[t]{0.13\textwidth}
			\centering
			\includegraphics[width=1\textwidth]{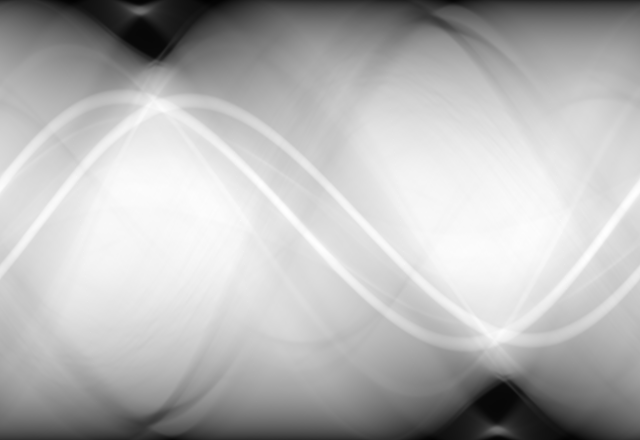}
			\includegraphics[width=1\textwidth]{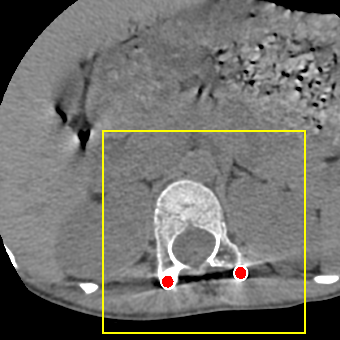}
			\includegraphics[width=1\textwidth]{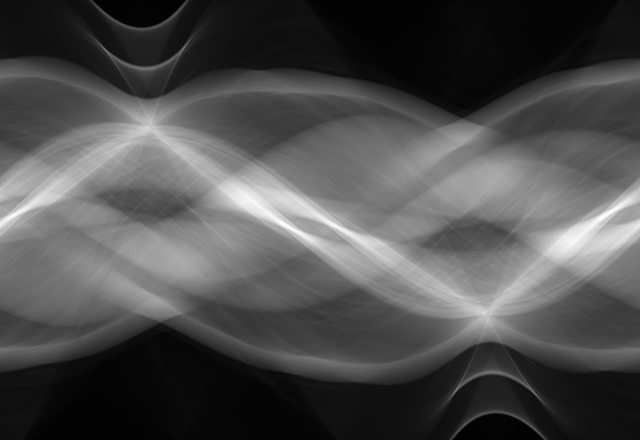}
			\includegraphics[width=1\textwidth]{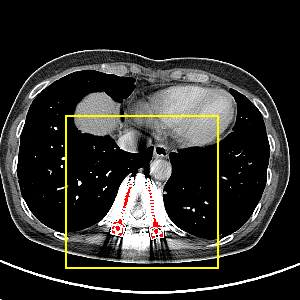}
			$S_{ma}/X_{ma}$
		\end{minipage}
		\begin{minipage}[t]{0.13\textwidth}
			\centering
			\includegraphics[width=1\textwidth]{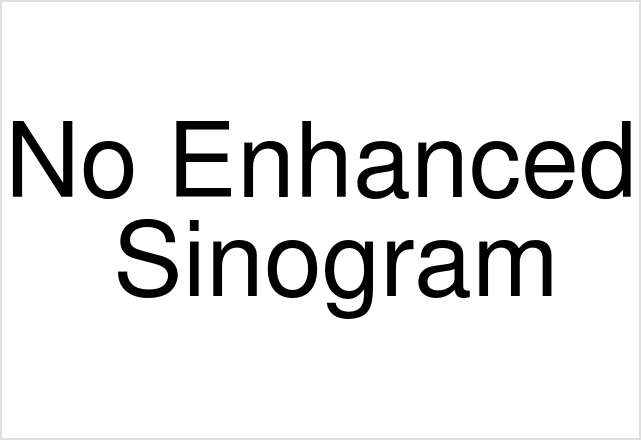}
			\includegraphics[width=1\textwidth]{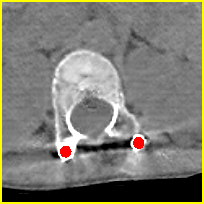}
			\includegraphics[width=1\textwidth]{figures/real_data/white.png}
			\includegraphics[width=1\textwidth]{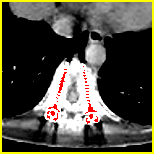}
			cGAN-CT
		\end{minipage}
		\begin{minipage}[t]{0.13\textwidth}
			\centering
			\includegraphics[width=1\textwidth]{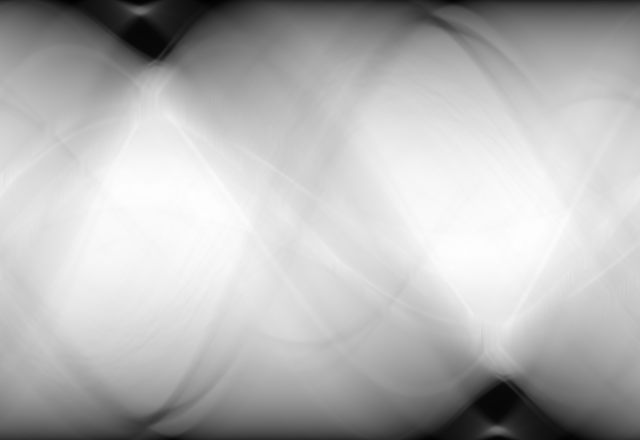}
			\includegraphics[width=1\textwidth]{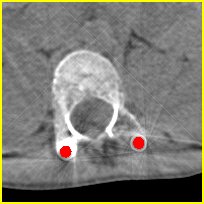}
			\includegraphics[width=1\textwidth]{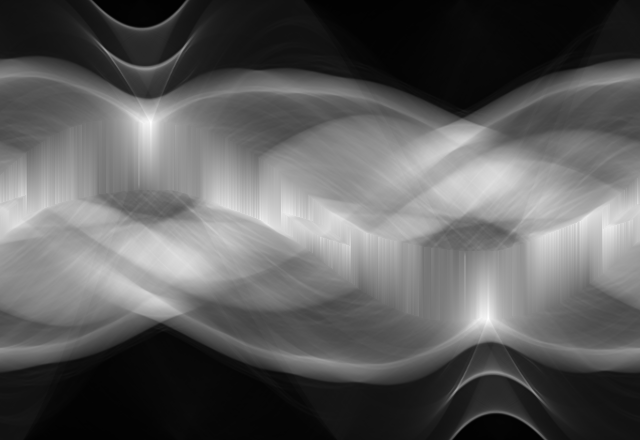}
			\includegraphics[width=1\textwidth]{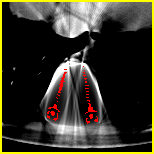}
			LI
		\end{minipage}
		\begin{minipage}[t]{0.13\textwidth}
			\centering
			\includegraphics[width=1\textwidth]{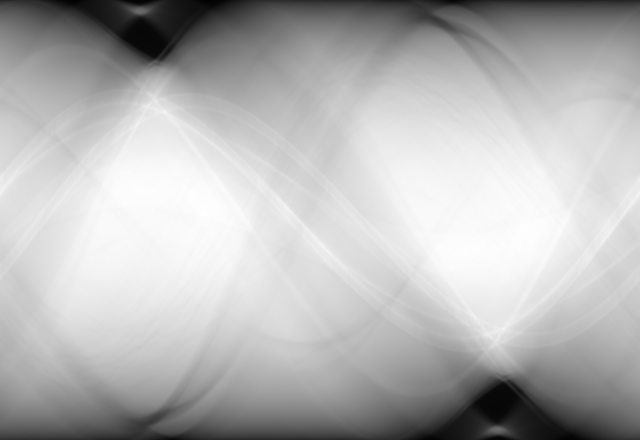}
			\includegraphics[width=1\textwidth]{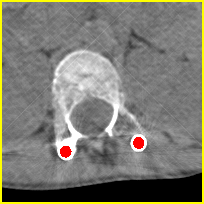}
			\includegraphics[width=1\textwidth]{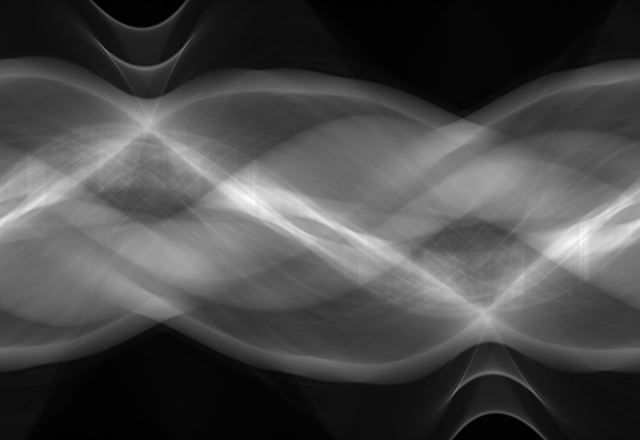}
			\includegraphics[width=1\textwidth]{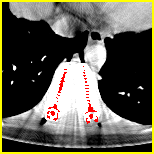}
			NMAR
		\end{minipage}
		\begin{minipage}[t]{0.13\textwidth}
			\centering
			\includegraphics[width=1\textwidth]{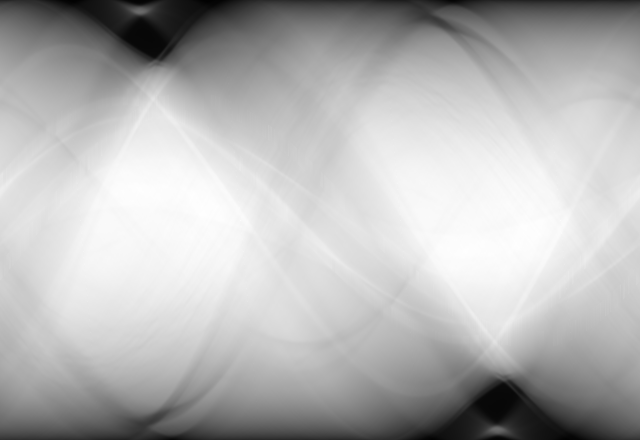}
			\includegraphics[width=1\textwidth]{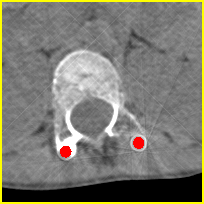}
			\includegraphics[width=1\textwidth]{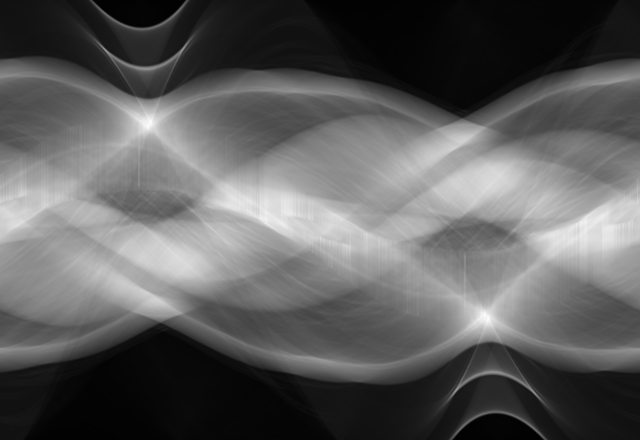}
			\includegraphics[width=1\textwidth]{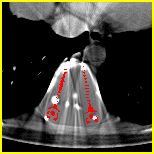}
			CNNMAR
		\end{minipage}
		\begin{minipage}[t]{0.13\textwidth}
			\centering
			\includegraphics[width=1\textwidth]{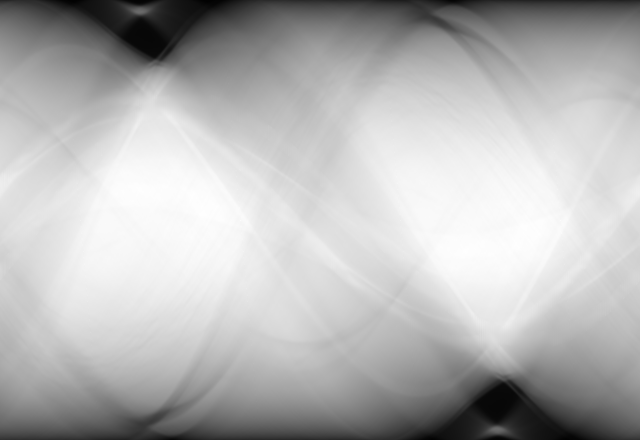}
			\includegraphics[width=1\textwidth]{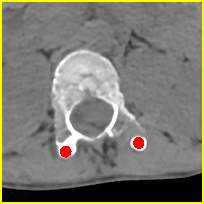}
			\includegraphics[width=1\textwidth]{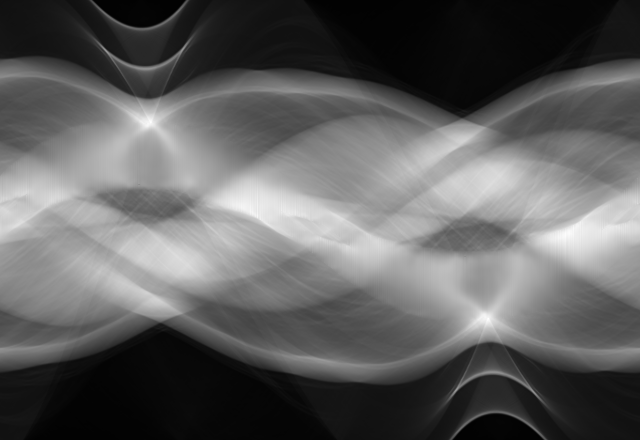}
			\includegraphics[width=1\textwidth]{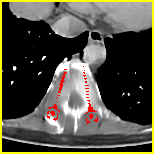}
			DuDoNet
		\end{minipage}
		\begin{minipage}[t]{0.13\textwidth}
			\centering
			\includegraphics[width=1\textwidth]{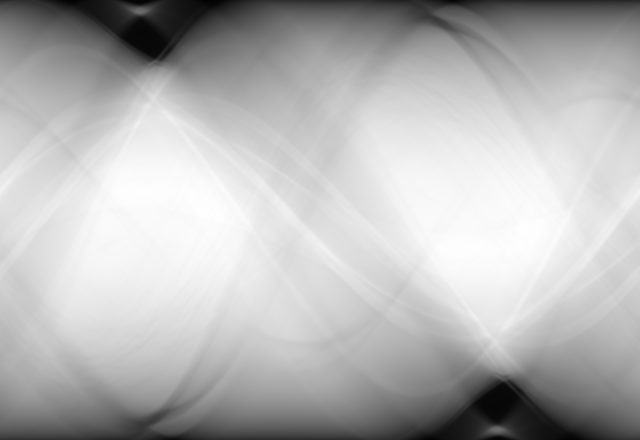}
			\includegraphics[width=1\textwidth]{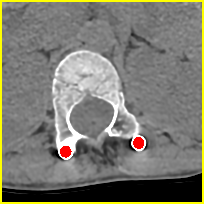}
			\includegraphics[width=1\textwidth]{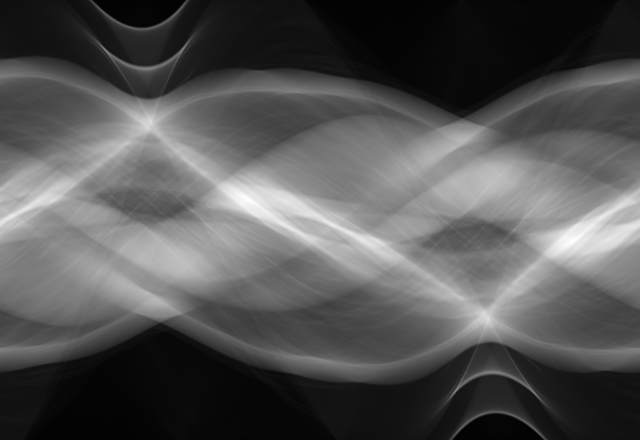}
			\includegraphics[width=1\textwidth]{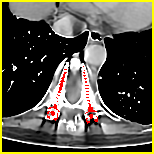}
			Ours
		\end{minipage}
		\caption{Comparison with the state-of-the-art methods on clinical CT images with metal artifacts.}
		\label{fig:real_data}
	\end{figure*}

	\noindent\textbf{Visual comparison.} Fig. \ref{fig:real_data} shows two clinical CT images with metal artifacts. Case 1 is with moderate metal artifacts. cGan-CT does not suppress the artifacts completely and generates some fake details. LI, NMAR, CNN-MAR remove all the artifacts but introduce new streak artifacts, which is caused by the discontinuity in the corrected sinogram. DuDoNet outputs over-smoothed sinogram, which leads to blurred tissues close to the metal implants, such as muscle and bone. Only our model can provide realistic enhanced sinogram and remove the artifacts while retaining the structure of nearby tissues. Case 2 is very challenging as the rods bring strong metal shadows and bright artifacts around the vertebra. cGan-CT recovers the shape of vertebra but changes the overall image intensity. Other sinogram inpainting methods fail as the soft tissue and bone near the rods are heavily distorted. Our model removes part of the dark bands and reproduces correct anatomical structures around the rods. 
	
	The results show that our model generalizes well for clinical images with unknown metal materials and geometries. We generate simulate training data using titanium and will retrain the model with multiple metal materials to make it more robust. Meanwhile, images with unknown geometry would be processed in the same simulation space. But it is worth noting that our model is limited to 2D geometry and the metal artifacts in 3D projection (e.g. cone-beam CT) are much more challenging.
	
	\section{Conclusion}
	We present a novel model to better solve the metal artifact reduction problem. We propose encoding mask projection for the sinogram restoration while utilizing the metal-affected real image and sinogram to retain the rich information in dual-domain learning. With the fine details recovered in metal trace, our model can correctly restore the underlying anatomical structure even with large metallic objects present. Visual comparisons and qualitative evaluations demonstrate that our model yields better image quality than competing methods and exhibits a great potential of reducing CT metal artifacts even when applied to clinical images. In the future, we plan to conduct a large scale clinical study to thoroughly evaluate our approach in real clinical practices.
	

	%

\end{document}